\documentclass[a4paper,11pt]{article}
\pdfoutput=1 
\usepackage{jcappub}
\usepackage{amsmath}
\usepackage{graphicx}
\usepackage{latexsym}
\usepackage{xspace}
\usepackage{color}
\usepackage{hyperref} 
\usepackage{bm}
\usepackage{relsize}
\usepackage{tabularx}
\usepackage{multirow}
\usepackage{amssymb}
\usepackage[table]{xcolor}
\usepackage{braket}
\usepackage{comment}
\usepackage[utf8]{inputenc}
\usepackage{slashed}
\usepackage{empheq}
\usepackage{cancel}

\usepackage[margin=1in,includefoot]{geometry}
\usepackage{xfrac} 
\usepackage[compat=1.1.0]{tikz-feynman}

\newcommand{\beq}{\begin{equation}}
\newcommand{\eeq}{\end{equation}}
\newcommand{\bea}{\begin{equation}\begin{aligned}}
\newcommand{\eea}{\end{aligned}\end{equation}}
\newcommand{\Mp}{M_{\mathrm{\footnotesize{Pl}}}}
\newcommand{\Sec}[1]{Sec.~\ref{#1}}

\newcommand{\Eq}[1]{eq.~(\ref{#1})}
\newcommand{\Eqs}[1]{eqs.~(\ref{#1})}
\newcommand{\dd}{\mathrm{d}}
\newcommand{\ie}{\textsl{i.e.}}
\newcommand{\efolds}{$e$-folds}

\newcommand{\uin}{\mathrm{in}}
\newcommand{\uend}{\mathrm{end}}
\newcommand{\uc}{\mathrm{c}}
\newcommand{\Pfpt}[1]{P_{\mathrm{\footnotesize{FPT}},#1}}
\newcommand{\fNL}{f_{\mathrm{\footnotesize{NL}}}}

\allowdisplaybreaks

\makeatletter
\gdef\@fpheader{}
\g@addto@macro\bfseries{\boldmath}
\makeatother

\title{Quantum diffusion and large primordial perturbations from inflation\\$ $\\
{\large Invited chapter to the book ``Primordial Black Holes'', Springer 2024,} \\
\vspace{-0.4cm}
{\small ed. Chris Byrnes, Gabriele Franciolini, Tomohiro Harada, Paolo Pani, Misao Sasaki.}}

\author[a]{Vincent Vennin,}
\affiliation[a]{Laboratoire de Physique de l'\'Ecole Normale Sup\'erieure, ENS, CNRS, Universit\'e PSL, Sorbonne Universit\'e, Universit\'e Paris Cit\'e, F-75005 Paris, France}
\emailAdd{vincent.vennin@ens.fr}

\author[b]{David Wands}
\affiliation[a]{Institute of Cosmology \& Gravitation, University of Portsmouth, Dennis Sciama Building, Burnaby Road, Portsmouth, PO1 3FX, United Kingdom}
\emailAdd{david.wands@port.ac.uk}

\date{today}

\begin{document}
	\sloppy
	
	\abstract{Quantum diffusion describes the inflow of vacuum quantum fluctuations as they get amplified by gravitational instability, and stretched to large distances during inflation. In this picture, the dynamics of the universe's expansion becomes stochastic, and the statistics of the curvature perturbation is encoded in the distribution of the duration of inflation. This provides a non-perturbative framework to study cosmological fluctuations during inflation, which is well-suited to the case of primordial black holes since they originate from large fluctuations. We show that standard, perturbative expectations for the primordial black hole abundance can be significantly modified by quantum-diffusion effects, and we identify a few open challenges.
	
		\vspace{1cm} \textit{In memory of Alexei Starobinsky (1948-2023), the founding father of stochastic inflation, and of so much more.}
}
	
	
	
	\maketitle

	\flushbottom

\section{Introduction}\label{sec:intro}

PBHs may form in the early universe when large density fluctuations are produced and gravitationally collapse into black holes. Although different mechanisms have been proposed to amplify cosmological fluctuations above the required threshold, one possibility is that the amplification takes place during inflation. If large curvature perturbations are produced during inflation, they may no longer be described by perturbative frameworks, and in this chapter we show how the stochastic-inflation formalism, in combination with the $\delta N$ formalism, can be used as a non-perturbative scheme to keep track of large fluctuations. In this approach, quantum fluctuations provide random kicks to the large-scale dynamics of the universe, which can thus be described by means of stochastic systems. Different regions of space inflate for different amounts of time, hence the number of inflationary $e$-folds becomes itself a stochastic quantity, which can be identified with the curvature perturbation. We show how its statistics can be studied using first-passage-time techniques, and discuss the expected consequences for the abundance of PBHs from inflation. This abundance is strongly enhanced compared to perturbative predictions due to heavy tails in the probability distribution, which are non-perturbative in nature and inevitable.  We also discuss the validity of the separate-universe approach, on which this framework rests, in particular in the context of slow-roll violations, which commonly arise in PBH-producing models. Finally, we identify a few open challenges that remain to be addressed. 

\section{Stochastic inflation in a nutshell}\label{sec:StochasticInflation}

The stochastic formalism of inflation~\cite{Starobinsky:1982ee, Starobinsky:1986fx} is an effective theory for the long-wavelength part of quantum fields living on an inflating background. It can be obtained formally by integrating out the small-wavelength part of these fields, using the language of effective-field theories and the Schwinger-Keldysh formalism~\cite{Morikawa:1989xz, Hu:1994dka, Matarrese:2003ye, Tokuda:2017fdh, Tokuda:2018eqs, Pinol:2020cdp}. Here, we present a heuristic derivation that makes physical interpretation easier. 

We consider a single scalar field $\phi$, minimally coupled to gravity in a 4-dimensional curved space-time with metric $g_{\mu\nu}$, described by the action
\bea
	S=\displaystyle\int \dd^4x\sqrt{-g}\left[\frac{\Mp^2}{2}R-\frac{1}{2}g^{\mu\nu}\partial_\mu\phi\partial_\nu\phi-V(\phi)\right] .
	\label{eq:action}
\eea
In this expression, $R$ is the Ricci scalar curvature and $V(\phi)$ is the potential energy of the scalar field. 

\subsection{Background}
At the background level, space-time is homogeneous and isotropic, so the line-element is of the Friedmann-Lema\^itre-Robertson-Walker (FLRW) form, $\dd s^2=-\dd t^2 + a^2(t) \dd x^2$. Here, $t$ is cosmic time, $a$ is the scale factor and $x^i$ are spatial coordinates comoving with the expansion. The Einstein equations yield a constraint equation, called the Friedmann equation,
\bea
\label{eq:Friedmann}
H^2 = \frac{V(\phi) + \frac{\dot{\phi}^2}{2}}{3\Mp^2}
\eea
where $H=\dot{a}/a$ is the Hubble expansion rate, a dot denotes a derivative with respect to cosmic time and $\Mp=\sqrt{\hbar c/(8\pi G)}$ is the reduced Planck mass. The dynamical equations for the inflaton field, $\phi$, can be written as two coupled first-order equations
\bea
\frac{\dd\phi}{\dd N} = &\frac{\Pi}{a^3 H}\, ,\\
\frac{\dd\Pi}{\dd N} =& -\frac{a^3}{H}V'(\phi)\, ,
\eea
where $\Pi$ is the momentum conjugate to $\phi$, $V'$ is the first derivative of the function $V(\phi)$, and time is labeled by the number of \efolds~$N\equiv\ln(a)$ for future convenience. When the potential function is flat enough, $\Mp \vert V'/V \vert\ll 1$ and $\Mp^2 \vert V''/V \vert\ll 1$, a phase of slow-roll inflation takes place where $\ddot{a}>0$ and $H$ varies slowly.

\subsection{Linear perturbations}
\label{sec:linearperturbations}

Let us now introduce small fluctuations over the homogeneous and isotropic background, by allowing $\phi$ to depend on space, $\phi(\vec{x},t)=\bar{\phi}(t) + \delta\phi(\vec{x},t)$, where $\bar{\phi}$ follows the background dynamics given above and $|\delta\phi|\ll|\bar{\phi}|$.\footnote{Metric perturbations are not included at this stage. The conditions under which this is a valid approximation when time is labeled with $N$ are clarified in \Sec{sec:SeparateUniverse}.} Inserting this decomposition into \Eq{eq:action}, at leading order in perturbation theory one has
\bea
\label{eq:eom:lin:deltaphi}
\frac{\dd\phi_ k}{\dd N} = &\frac{\pi_k}{a^3 H}\, ,\\
\frac{\dd\pi_k}{\dd N} =& -\frac{a^3}{H}V''(\bar{\phi})\phi_k-\frac{a}{H}k^2\phi_k\, .
\eea
These equations are written in Fourier space, \ie
\bea
\delta\phi(\vec{x},t) = \left(2\pi\right)^{-3/2}\int\dd\vec{k} \left[ a_{\vec{k}} \phi_{k}(t) e^{-i\vec{k}\cdot\vec{x}}+ a_{\vec{k}}^\dagger \phi^*_{k}(t) e^{i\vec{k}\cdot\vec{x}}\right]
\eea
with a similar expression for $\delta\pi(\vec{x},t)$. Here, the field $\delta\phi(\vec{x},t)$ is quantised, so it is expanded onto the creation and annihilation operators $a^\dagger_{\vec{k}}$ and $a_{\vec{k}}$, which satisfy the usual commutation relations $[a_{\vec{k}},a^\dag_{\vec{k}'}]=\delta^3(\vec{k}-\vec{k}')$ and $[a_{\vec{k}},a_{\vec{k}'}]=[a^\dag_{\vec{k}},a^\dag_{\vec{k}'}]=0$. They are independent of time, contrary to the mode functions $\phi_k(t)$ which satisfy \Eqs{eq:eom:lin:deltaphi}. On isotropic backgrounds, they depend only on the modulus $k=\vert\vec{k}\vert$ of the wave-vector, and they must satisfy $\phi_{-\vec{k}}=\phi_{\vec{k}}^\star$ for the quantum field $\delta\phi(\vec{x},t)$ to be Hermitian (\ie~for the classical field to be real). In the asymptotic past, $k\gg aH$ and the above reduces to the dynamics of a massless field on a Minkowski background. In this sub-Hubble regime, each Fourier mode is described by a harmonic oscillator, the ground state of which can serve as an initial condition. This is the so-called Bunch-Davies vacuum, which corresponds to setting $\phi_k=e^{-ik\eta}/(a\sqrt{2k})$ and $\pi_k=-ika^2\phi_k$ as $k\eta\to-\infty$, where $\eta$ is the conformal time, related to the cosmic time $t$ via $\dd t= a\dd\eta$. Together with \Eq{eq:eom:lin:deltaphi}, this specifies the state of the perturbations at any time. In particular, this is a Gaussian state, given that the ground state of the harmonic oscillator is Gaussian and that the evolution is linear. Non-Gaussianities arise at non-linear orders, and one of our goals is to describe them non-perturbatively.

\subsection{Coarse-grained fields}
\label{sec:coarsegrain}

In practice, cosmological perturbation theory provides an accurate description of cosmological fluctuations so long as their amplitude remains small. In the context of PBHs, one is precisely interested in situations where the rare, high peaks of cosmological perturbations grow on super-Hubble scales and become sizeable, hence non-perturbative frameworks are required at large scales. One such framework is the stochastic-inflation formalism, which describes the dynamics of fields coarse-grained at the $\sigma$-Hubble radius, $(\sigma H)^{-1}$. Here, $\sigma$ is a fixed parameter that sets the scale at which quantum fluctuations backreact onto the local FLRW geometry. In practice, one must set $\sigma\ll 1$ in order for gradient terms not to affect the dynamics of the coarse-grained field, see \Sec{sec:SeparateUniverse}. One thus decomposes
\bea
\label{eq:field;decomposition:phi}
\phi=\tilde{\phi}+\phi_Q\\
\Pi=\tilde{\pi}+\pi_Q
\eea
where $\tilde{\phi}$ and $\tilde{\pi}$ are coarse-grained fields, and 
\bea
\label{eq:phiq}
	\phi_Q&=&\displaystyle\int_{\mathbb{R}^3}\frac{\dd^3 \vec{k}}{(2\pi)^{3/2}}W\left(\frac{k}{k_\sigma}\right)\left[a_{\vec{k}}~\phi_{k}(\tau)e^{-i\vec{k}\cdot\vec{x}}+a^\dag_{\vec{k}}~\phi^\star_{k}(\tau)e^{i\vec{k}\cdot\vec{x}}\right] \\
	\pi_Q&=&\displaystyle\int_{\mathbb{R}^3}\frac{\dd^3 \vec{k}}{(2\pi)^{3/2}}W\left(\frac{k}{k_\sigma}\right)\left[a_{\vec{k}}~\pi_{k}(\tau)e^{-i\vec{k}\cdot\vec{x}}+a^\dag_{\vec{k}}~\pi^\star_{k}(\tau)e^{i\vec{k}\cdot\vec{x}}\right]
\eea
are their small-wavelength complements. Here, $k_\sigma=\sigma aH$, and the window function $W$ is such that $W\simeq0$ for $k \ll k_\sigma$ and $W\simeq1$ for $k \gg k_\sigma$. The equations of motion for the coarse-grained fields are given by
\bea
\label{eq:Hamilton:split:phi:pi}
	\frac{\dd{\tilde\phi}}{\dd N}=&\frac{\tilde\pi}{a^3H}-\frac{\dd\phi_Q}{\dd N}+\frac{\pi_Q}{H a^{3}}\,  , \\
	\frac{\dd{\tilde\pi}}{\dd N}=&-\frac{a^{3}}{H}V'(\tilde\phi)-\frac{\dd\pi_Q}{\dd N}-\frac{a^{3}}{H}V''(\tilde\phi)\phi_Q+\frac{a}{H} \Delta\phi_Q .
\eea
In these expressions, the Laplacian of $\tilde\phi$ has been dropped since it is suppressed by $\sigma$, and a linear expansion in the small-wavelength perturbations $\phi_Q$ and $\pi_Q$ has been performed. Replacing $\phi_Q$ and $\pi_Q$ by \Eqs{eq:phiq}, and making use of the fact that the mode functions $\phi_k$ and $\pi_k$ satisfy \Eqs{eq:eom:lin:deltaphi}, the above can be rewritten as~\cite{Nakao:1988yi, Habib:1992ci, Rigopoulos:2005xx, Tolley:2008na, Weenink:2011dd, Grain:2017dqa}
\bea
\label{eq:Langevin}
	\frac{\dd{\tilde\phi}}{\dd N}=&\frac{\tilde\pi}{a^3H}+\xi_\phi(N)\,  , \\
	\frac{\dd{\tilde\pi}}{\dd N}=&-\frac{a^{3}}{H}V'(\tilde\phi)+\xi_\pi(N)\, ,
\eea
where the quantum sources $\xi_\phi$ and $\xi_\pi$ arise from the time dependence of $k_\sigma$ in the window functions in \Eq{eq:phiq} and are given by
\bea
 \label{eq:noisephipi}
	\xi_\phi&=&-\displaystyle\int_{\mathbb{R}^3}\frac{\dd^3\vec{k}}{(2\pi)^{3/2}}\frac{\dd{W}}{\dd N}\left(\frac{k}{k_\sigma}\right)\left[a_{\vec{k}}\phi_k(\tau)e^{-i\vec{k}\cdot\vec{x}}+a^\dag_{\vec{k}}\phi^\star_k(\tau)e^{i\vec{k}\cdot\vec{x}}\right] ,\\
	\xi_\pi&=&-\displaystyle\int_{\mathbb{R}^3}\frac{\dd^3\vec{k}}{(2\pi)^{3/2}}\frac{\dd{W}}{\dd N}\left(\frac{k}{k_\sigma}\right)\left[a_{\vec{k}}\pi_k(\tau)e^{-i\vec{k}\cdot\vec{x}}+a^\dag_{\vec{k}}\pi^\star_k(\tau)e^{i\vec{k}\cdot\vec{x}}\right] . 
\eea
They act as source terms in the equations of the motion for the coarse-grained fields. Since they only involve fluctuations at the $\sigma$-Hubble scale, they can be evaluated in cosmological perturbation theory where they obey Gaussian statistics. Their properties are thus entirely captured by their two-point functions,
\bea
\label{eq:noise:corr}
\left\langle \xi_f(\vec{x}_1,N) \xi_g(\vec{x}_2,N')\right\rangle = \frac{\dd\ln k_\sigma}{\dd N} \mathcal{P}_{fg}\left[k_\sigma(N),N\right]\mathrm{sinc}\left[k_\sigma(N)\vert \vec{x}_2-\vec{x}_1\vert\right]\delta\left(N-N'\right)
\eea
where $f$ and $g$ denote either $\phi$ or $\pi$. Here, we have assumed that the window function is sharp in Fourier space, $W(k/k_\sigma)=\Theta(k/k_\sigma-1)$. This explains why the quantum noises are white, \ie~uncorrelated over time. In the above expression, we have also introduced the reduced power spectrum,
\bea 
\mathcal{P}_{fg}(k,N)= \frac{k^3}{2\pi^2} f_k(N) g_k^\star(N)\, .
\eea 

\subsection{Langevin equations}

The next step involves describing spatial hypersurfaces of the universe as ensembles of $\sigma$-Hubble patches. Between two such patches, the correlations between the quantum sources are suppressed by the cardinal sine function appearing in \Eq{eq:noise:corr}. Since gradient interactions have been already dropped when discarding the Laplacian of $\tilde{\phi}$ in \Eq{eq:Hamilton:split:phi:pi}, such correlations must be neglected at the order at which the calculation is performed. This implies that $\sigma$-Hubble patches evolve independently, which is usually referred to as the ``separate-universe'' picture. 

Before elaborating more on the validity of this assumption in \Sec{sec:SeparateUniverse}, let us note that, in this approach, each patch is endowed with a local FLRW geometry, in which the (local) expansion rate is still given by the Friedmann equation~(\ref{eq:Friedmann}), where the right-hand side must now be evaluated with the (local) values of $\tilde{\phi}$ and $\tilde{\pi}$. Such an equation may however seem problematic since, although $H$ is a geometrical classical quantity, at this stage $\tilde{\phi}$ and $\tilde{\pi}$ are still quantum fields. The present approach thus requires $\tilde{\phi}$ and $\tilde{\pi}$ to be seen as classical, random fields, rather than quantum fields. The quantum sources $\xi_\phi$ and $\xi_\pi$ become stochastic noises, with statistical correlation functions that are identified with the quantum expectation values~(\ref{eq:noise:corr}). 

Since $\xi_\phi$ and $\xi_\pi$ are Hermitian quantum fields placed in a Gaussian state, and given that they enter linearly in the equations of motion~(\ref{eq:Langevin}), they can be equivalently described by classical random variables sharing the same two-point functions, and the two descriptions are perfectly equivalent~\cite{2014arXiv1401.4679A, Martin:2015qta, Grain:2017dqa}. In contrast, the coarse-grained fields $\tilde{\phi}$ and $\tilde{\pi}$ may be placed in non-Gaussian states as an effect of non-linear evolution, and in this case the stochastic and quantum descriptions are not equivalent. It is indeed well-known that quantum systems can display correlations that cannot be accounted for using classical setups, a famous example being the celebrated Bell inequalities~\cite{Bell:1964kc}. In fact, even if $\tilde{\phi}$ and $\tilde{\pi}$ remain in Gaussian states, non-classical correlations appear when considering ``improper'' combinations,\footnote{In this context, ``improper'' denotes operators for which the Wigner-Weyl transform takes values outside their spectrum~\cite{2006FoPh...36..546R}.} for which Bell inequalities can be explicitly violated~\cite{Martin:2016tbd, Ando:2020kdz}. In practice however, these quantum signatures are often hidden in the amplitude of the so-called decaying mode~\cite{Polarski:1995jg, Lesgourgues:1996jc, Kiefer:2008ku, Martin:2017zxs}, which is again gradient suppressed, and may be further concealed by quantum decoherence~\cite{Martin:2021znx, Martin:2021qkg, Burgess:2022nwu, Espinosa-Portales:2022yok}. 

When such a ``quantum-to-classical'' transition takes place, the stochastic approach is valid, and \Eqs{eq:Langevin} may be seen as Langevin equations~\cite{risken1996fokker}. The white Gaussian noises $\xi_\phi$ and $\xi_\pi$ model the inflow of sub-Hubble quantum fluctuations into the super-Hubble, coarse-grained sector of the theory, which behaves as classical random fields.

\section{Inhomogeneous spacetime}\label{sec:SeparateUniverse}

Thus far we have considered the evolution of inhomogeneous scalar field fluctuations in an unperturbed FLRW spacetime. However, since it is the energy density of the scalar field that determines the Hubble expansion during inflation, see \Eq{eq:Friedmann}, then we should also consider inhomogeneous perturbations of the spacetime, consistently perturbing the metric as well as the field content. 
Traditionally when studying the evolution of inhomogeneous fluctuations in cosmology,
one allows for arbitrary, but small perturbations about the homogeneous and isotropic background and then systematically expands order-by-order in a small parameter controlling the deviation from FLRW symmetries~\cite{Malik:2008im}. However this perturbative approach is liable to break down where fluctuations become large, either due to the accumulation of many small perturbations over a long time period, e.g., due to quantum diffusion, or a sudden large enhancement in the variance of those perturbations at a particular time or scale, for example, due to a localised feature in the scalar field potential. 

In practice the coarse-graining scale, $k_\sigma$ introduced in section~\ref{sec:coarsegrain}, allows us to adopt a different non-perturbative approach to model the evolution of potentially large inhomogeneities on large scales, $k<k_\sigma$, while retaining a perturbative approach to quantum fluctuations on small scales. This is known as the separate universe approach~\cite{Sasaki:1998ug, Wands:2000dp} and it is an implicit, but essential part of the stochastic framework outlined above. It then allows us to reconstruct the inhomogeneous spacetime metric at the end of a period of stochastic inflation in terms of the perturbation of the local integrated expansion, $\delta N$~\cite{Sasaki:1998ug,Lyth:2003im,Lyth:2004gb}.

\subsection{Separate-universe approach}

An important simplification occurs when studying inhomogeneous perturbations on sufficiently large scales where anisotropy and spatial gradients can be neglected. In that case the evolution of the fields locally can be described using ordinary differential equations for the time-dependence of effectively homogeneous and isotropic fields in an FLRW cosmology, considerably simplifying the calculation, and enabling us to track the nonlinear as well as linear evolution. This is the separate-universe approach~\cite{Sasaki:1998ug, Wands:2000dp}. Formally we can interpret this as the leading-order behaviour in a gradient expansion~\cite{Salopek:1990jq, Rigopoulos:2003ak, Tanaka:2007gh, Launay:2024qsm}. 

For example, if we consider linear perturbations about an FLRW cosmology of the scalar field 
(as already introduced in subsection~\ref{sec:linearperturbations}) and in addition perturbing the lapse function (allowing for local dilation of the proper time, $\tau$ relative to the background cosmic time $t$):
\begin{equation}
\phi \to \bar\phi + \delta\phi \,, \quad \partial/\partial t \to \partial/\partial \tau = (1-A)\partial/\partial t  \, , 
\end{equation}
then, at first order in the perturbations, the Klein-Gordon equation for the scalar field at each point can be written as~\cite{Pattison:2019hef}
\begin{equation}
\label{eq:pertKG}
\frac{\partial^2}{\partial \tau^2}(\bar\phi+\delta\phi)+3(\bar{H}+\delta H)\frac{\partial}{\partial \tau}(\bar\phi+\delta\phi)+V_{,\phi}(\bar\phi+\delta\phi) = \nabla^2(\delta\phi) \,,
\end{equation}
where $\bar{H}+\delta H$ is the local expansion rate. Up to the spatial gradient term, $\nabla^2\delta\phi$, \Eq{eq:pertKG} has exactly the same form as the Klein-Gordon equation for a homogeneous scalar field $\tilde\phi=\bar\phi+\delta\phi$ in an FLRW spacetime with Hubble rate $H=\bar{H}+\delta H$. 
The separate-universe approach is expected be valid on scales larger than the Hubble radius, $H^{-1}$, such that spatial gradients of the fields are small relative to the expansion rate. 
Thus we identify $\tilde\phi$ and ${H}$ appearing in the Langevin equations (\ref{eq:Langevin}) with the field and expansion rate coarse-grained on scales larger than $(\sigma H)^{-1}$ where $\sigma\ll1$, and $H$ is given by the local Friedmann equation
\bea
\label{eq:Friedmann}
H^2 = \frac{1}{3\Mp^2} \left[ V(\tilde\phi) + \frac12 \left( \frac{\dd \tilde\phi}{\dd \tau}\right)^2 \right] \,.
\eea
The separate universe approach turns out to be particularly powerful during inflation, since the accelerated expansion of the universe ensures that the spatial gradients of comoving wavemodes decrease more quickly than the Hubble rate ($H>|\dot{H}/H|$). Thus spatial gradients rapidly decrease as modes are stretched beyond the Hubble scale. 

Intuitively we consider the coarse-grained field in each super-Hubble-sized patch of the universe to be evolving like a separate FLRW universe, where the fields may take different values in different patches, but are treated as homogeneous within each patch. After sewing back together the different patches, we can reconstruct the inhomogeneous evolution of our entire observable universe above the coarse-graining scale. To relate local quantities to a global coordinate system amounts to a choice of gauge. This cannot be determined by the local FLRW quantities but requires the use of additional constraint equations from cosmological perturbation theory. For example in the spatially-flat gauge the momentum constraint imposes a simple relation between the lapse function and the scalar-field perturbation~\cite{Kodama:1984ziu,Malik:2008im}
\[
A = \frac{\dot\phi}{2\Mp^2H}\delta\phi \,.
\]
In the slow-roll limit $\dot\phi\to0$ and the local proper time coincides with the background cosmic time, but beyond slow roll one must consistently account for local variations in the proper time interval in different patches to be able to relate the separate-universe evolution locally to the evolution of inhomogeneous perturbations about the global background~\cite{Pattison:2019hef}. 

Since the Langevin equations~(\ref{eq:Langevin}) use the logarithmic expansion, $N$, to describe the evolution of the scalar field and its canonical momentum in local patches,  when we describe the ensemble of many different patches at a given $N$, we are implictly using a gauge in which the integrated local expansion 
\begin{equation}
\label{eq:N}
N\equiv\int(\bar{H}+\delta H)\,(1+A)\,\dd t \,,
\end{equation}
remains unperturbed. 
Quantum field fluctuations crossing outside the Hubble scale during inflation are commonly calculated using the Sasaki-Mukhanov~\cite{Sasaki:1986hm,Mukhanov:1988jd} mode equation for scalar field fluctuations in the spatially-flat gauge~\cite{Kodama:1984ziu,Malik:2008im}. In slow-roll inflation the integrated local expansion is unperturbed in the spatially-flat gauge, but in general there is a gauge transformation required to consistently describe scalar field fluctuations in a uniform-$N$ gauge~\cite{Pattison:2019hef}.

While the separate universe approach works well on super-Hubble scales in many situations, it has recently been shown~\cite{Jackson:2023obv}  that it breaks down on a finite range of super-Hubble scales in inflationary models with features, such as an inflection point, leading to a sudden, non-adiabatic transition from slow roll to a transient, ultra-slow-roll phase. This boosts the amplitude of density perturbations on small scales, which exit the Hubble-scale after the transition, which could lead to the formation of primordial black holes after inflation, or a stochastic background of gravitational waves induced at second-order in perturbation theory. But sudden transitions can also lead to spatial gradients generating non-adiabatic perturbations on super-Hubble scales at the transition~\cite{Leach:2001zf}, breaking the usual assumption that the separate universe approach can safely be used soon after Hubble exit. In such a situation one should be cautious about applying the usual stochastic-inflation formalism, for example, by waiting until after a sudden transition~\cite{Jackson:2023obv}, or until the transient, non-adiabatic perturbations have decayed, before using the Langevin equations~\cite{Tomberg:2023kli}.

\subsection{$\delta N$ formalism}

Primordial density perturbations at the end of inflation are commonly characterised in terms of the gauge-invariant variable, $\zeta$, which corresponds to the scalar metric perturbation, $-\psi$, on uniform-density hypersurfaces ($\delta\rho=0$). At first-order this is written in terms of the curvature and density perturbations in an arbitrary gauge as~\cite{Malik:2008im}
\begin{equation}
\label{eq:zetalinear}
\zeta \equiv -\psi - \frac{H\delta\rho}{\dot\rho} \,.
\end{equation}
In particular it can be identified with the dimensionless density perturbation on spatially-flat hypersurfaces
\begin{equation}
\label{eq:zeta-alt}
\zeta = - \left. \frac{H\delta\rho}{\dot\rho} \right|_{\psi=0} \,.
\end{equation}
It can be shown that $\zeta$ is conserved for adiabatic perturbations, irrespective of the field or matter content of the universe, in the large-scale limit where spatial gradients can be neglected~\cite{Wands:2000dp,Weinberg:2003sw}. Hence it is a particularly useful tool to link scalar-field fluctuations during inflation to perturbations in the primordial plasma some time after inflation~\cite{Bardeen:1983qw}. 

More generally $\zeta$ can be identified with the nonlinearly perturbed expansion between a spatially-flat hypersurface ($\psi=0$) and the uniform-density hypersurface~\cite{Sasaki:1998ug,Lyth:2003im,Lyth:2004gb}
\begin{equation}
\label{eq:zetanonlinear}
\zeta = \delta N \equiv N\left[\vec{\Phi}(\vec{x})\right] - \langle N(\vec{\Phi}) \rangle \,,
\end{equation}
where the local expansion $N$ was defined in equation~(\ref{eq:N}). In our definition of $\zeta$ in (\ref{eq:zetanonlinear}) we require that the lower limit of the integral ~(\ref{eq:N}) is a spatially-flat hypersurface, where the initial field-phase space vector at some point is given by $\vec{\Phi}(\vec{x})$, and the upper limit of the integral is a uniform-density hypersurface at the end of inflation corresponding to the field-phase space vector $\vec{\Phi}_{\mathrm{end}}$ at the end of inflation.

Thus we are able to calculate the distribution of primordial density perturbations after inflation in terms of the distribution of values of the perturbed local expansion, $N$, from an initial spatially-flat hypersurface during inflation to a uniform-density hypersurface at the end of inflation. If we employ the separate-universe approach to determine the nonlinear evolution on large scales, then we can use the background FLRW equations of motion to determine local expansion, $N(\vec{\Phi})$, as a function of the local coarse-grained field-phase space values $\vec{\Phi}(\vec{x})$~\cite{Starobinsky:1982ee,Starobinsky:1985ibc,Sasaki:1995aw,Lyth:2005fi}. 

In what we will call the {\em classical} $\delta N$ formalism the fields are coarse-grained on one initial hypersurface such that all comoving scales of interest are above the Hubble scale at that time. If we further restrict ourselves to slow-roll evolution where the local field momenta are a function the local field values, we can then formally expand \Eq{eq:zetanonlinear} to give~\cite{Lyth:2005fi}
\begin{equation}
\label{eq:zetanonlinear:2}
\zeta = \sum_I \frac{\partial N}{\partial\phi^I} \delta\phi_I + \frac12 \sum_{I,J} \frac{\partial^2 N}{\partial\phi^I\partial\phi^I} \delta\phi_I \delta\phi_J + \ldots \,.
\end{equation}
Furthermore for single-field slow-roll inflation we have $\partial N/\partial\phi=-H/\dot\phi$ and hence \Eq{eq:zetanonlinear:2} reduces to \Eq{eq:zeta-alt} at first order, where $\rho=\rho(\phi)$.

The classical $\delta N$ formalism assumes that the subsequent evolution above one fixed comoving coarse-graining scale is independent of further fluctuations on smaller scales and can be described by the unperturbed background equations. In the {\em stochastic} $\delta N$ formalism~\cite{Vennin:2015hra} we choose a time-dependent coarse-graining scale, determined by the physical Hubble scale at any given time, thus incorporating the effect of quantum diffusion due to fluctuations on all scales that cross the coarse-graining scale before the end of inflation. The integrated expansion up to the end of inflation becomes a stochastic variable, ${\cal N}[\vec{\Phi}(\vec{x})]$, whose statistical properties are a function of the initial field-phase space vector.

\section{First-passage-time analysis}\label{sec:FPT}

The Langevin equations~(\ref{eq:Langevin}) can be written more generally as
\bea
\label{eq:Lang:gen}
\frac{\dd{\Phi}^i}{\dd N} = F^i(\vec{\Phi}) + G^i_{j}(\vec{\Phi})\xi^j(N)\, ,
\eea
where $\vec{\Phi}$ is the field-phase space vector that comprises all fields and conjugate momenta. The classical, homogeneous equations of motion are contained in $\vec{F}$, while $\vec{G}$ is obtained by evolving the field fluctuations from the Bunch-Davies vacuum to the $\sigma$-Hubble crossing time. The white Gaussian noises $\xi_i$ are normalised such that $\langle \xi^i(N) \xi^j(N')\rangle = \delta^{i,j}\delta(N-N')$. In the case of single-field inflation with a canonical kinetic term, $\vec{F}$ and $\vec{G}$ can be read off \Eqs{eq:Langevin}. Along the slow-roll attractor, quantum diffusion only occurs along the classical flow lines~\cite{Grain:2017dqa}, hence field phase space reduces to $\vec{\Phi}=(\phi)$ and one has $F=-V'/(3H^2)$ and $G=H/(2\pi)$, with $H^2=V/(3\Mp^2)$. For now, let us keep the generic form~(\ref{eq:Lang:gen}), since it allows us to describe any setup, including multi-fields models in non-trivial field-space geometries~\cite{Pinol:2020cdp}. 

\subsection{Fokker-Planck equation}

From the Langevin equation~\eqref{eq:Lang:gen}, one can derive a Fokker-Planck equation that drives the probability $P(\vec{\Phi},N \vert \vec{\Phi}_\uin,N_\uin)$ to find the configuration $\vec{\Phi}$ at time $N$ knowing that the system was initiated as $\vec{\Phi}_\uin$ at time $N_\uin$. It is given by~\cite{risken1996fokker}
\bea
\label{eq:Fokker:Planck}
\frac{\partial }{\partial N}P(\vec{\Phi},N \vert \vec{\Phi}_\uin,N_\uin) = &\left(-\frac{\partial}{\partial\Phi^i} F^i+ \frac{1}{2} \frac{\partial^2}{\partial\Phi^i \partial \Phi^j} G^j_k G^{ki}\right)P(\vec{\Phi},N \vert \vec{\Phi}_\uin,N_\uin)\\ 
\equiv & \mathcal{L}(\vec{\Phi})\cdot P(\vec{\Phi},N \vert \vec{\Phi}_\uin,N_\uin) \, ,
\eea
where we use It\^o's discretisation convention. This defines the Fokker-Planck operator $\mathcal{L}(\vec{\Phi})$, which is a differential operator acting in the $\vec{\Phi}$ direction. The Fokker-Planck equation needs to be solved from the initial condition $P(\vec{\Phi},N_\uin \vert \vec{\Phi}_\uin,N_\uin)=\delta(\vec{\Phi}-\vec{\Phi}_\uin)$, and with an absorbing boundary condition on the end-of-inflation hypersurface $\mathcal{C}_\uend$, such that $P(\vec{\Phi},N \vert \vec{\Phi}_\uin,N_\uin)=0$ when $\vec{\Phi}\in \mathcal{C}_\uend$, where $\mathcal{C}_\uend$ is of uniform energy density. 

The Langevin equation~\eqref{eq:Lang:gen} describes a Markovian process, hence for any fixed intermediate time $\bar{N}$, using the rules of conditional probabilities one has
\bea
P\left(\vec{\Phi},N \vert \vec{\Phi}_\uin,N_\uin\right) = \int_\Omega \dd \bar{\vec{\Phi}} P\left(\vec{\Phi},N \vert \bar{\vec{\Phi}},\bar{N}\right) P\left(\bar{\vec{\Phi}},\bar{N} \vert \vec{\Phi}_\uin,N_\uin\right) ,
\eea
where $\Omega$ is the inflating domain of field-phase space, bounded by $\mathcal{C}_\uend$. By differentiating both sides of this relation with respect to $\bar{N}$, one obtains
\bea
\label{eq:FP:adjoint}
\frac{\partial}{\partial \bar{N}} P\left(\vec{\Phi},N \vert \bar{\vec{\Phi}},\bar{N}\right)= \mathcal{L}^\dagger(\bar{\vec{\Phi}}) \cdot P\left(\vec{\Phi},N \vert \bar{\vec{\Phi}},\bar{N}\right) ,
\eea
where $\mathcal{L}^\dagger$ is adjoint to the Fokker-Planck operator $\mathcal{L}$ in the sense that, for any two field-phase space functions $f(\vec{\Phi})$ and $g(\vec{\Phi})$, $\int  \mathrm{d}\vec{\Phi}f(\vec{\Phi}) [ \mathcal{L}(\vec{\Phi}) \cdot  g(\vec{\Phi}) ] = \int \mathrm{d}\vec{\Phi} [ \mathcal{L}^\dagger(\vec{\Phi}) \cdot  f(\vec{\Phi})]  g(\vec{\Phi})$.  This is why \Eq{eq:FP:adjoint} is called the adjoint Fokker-Planck equation, or Kolmogorov backward equation, where, using integration by parts
\bea
 \mathcal{L}^\dagger(\vec{\Phi})  = F^i \frac{\partial}{\partial\Phi^i} + \frac{1}{2}G^i_k G^{ki} \frac{\partial^2}{\partial\Phi^i \partial\Phi^j}\, .
\eea

Our goal is to reconstruct the distribution of first-passage times $\mathcal{N}$ through the end-of-inflation hypersurface $\mathcal{C}_\uend$, starting from a given field configuration $\vec{\Phi}_\uin$, and identify it with the coarse-grained curvature perturbation using \Eq{eq:zetanonlinear}. This is the so-called stochastic-$\delta N$ program~\cite{Enqvist:2008kt, Fujita:2013cna, Vennin:2015hra, Pattison:2017mbe}. The first-passage time distribution is denoted $\Pfpt{\vec{\Phi}_\uin}(\mathcal{N})$, and is a central topic in stochastic analysis, see for instance Ref.~\cite{Redner_2001}. Here, we only give a few key results, following Refs.~\cite{Vennin:2015hra, Pattison:2017mbe, Ezquiaga:2019ftu} in the context of inflationary cosmology. 

A direct way to reconstruct the distribution of first passage times is through numerical sampling of the Langevin equation~\eqref{eq:Langevin}: for each realisation, record the duration of inflation $\mathcal{N}$, and draw a histogram (or use more advanced kernel reconstruction methods). This approach is numerically expensive, since a large number of realisations needs to be simulated to reach satisfactory levels of statistical noise, especially in the tail of the first-passage-time distribution where the statistics is scarce. Importance sampling can be used to circumvent this issue, see Ref.~\cite{Jackson:2022unc}, or methods based on the formalism of stochastic excursions, see Ref.~\cite{Tokeshi:2023swe}. However, for analytical insight it is useful to derive a driving equation for $\Pfpt{\vec{\Phi}_\uin}(\mathcal{N})$, analogous to the Fokker-Planck equation for $P(\vec{\Phi},N \vert \vec{\Phi}_\uin,N_\uin)$. This is the goal of this section.

\subsection{A warm up: free diffusion}
\label{eq:free:diffusion}

Let us start with the simple toy model consisting of a scalar field $\phi$ diffusing along a constant potential, along the slow-roll attractor. Upon introducing the rescaled variable $x=2\pi (\phi-\phi_\uend)/H$, the Langevin equation reads $\dd x/\dd N = \xi(N)$, and the Fokker-Planck equation is given by
\bea
\label{eq:free:diff:FP}
\frac{\partial }{\partial N}P(x,N \vert x_\uin, N_\uin) = \frac{1}{2}\frac{\partial^2}{\partial x^2} P(x,N \vert x_\uin, N_\uin)\, ,
\eea
which admits Gaussian solutions of the form
\bea
f_\mathrm{x_0,\sigma_0}(x,N) = \frac{e^{-\frac{(x-x_0)^2}{2\sigma^2(N)}}}{\sqrt{2\pi \sigma^2(N)}} \quad\quad \text{with} \quad\quad \sigma^2(N) =  \sigma_0^2 + N-N_\uin\, .
\eea
Since it is linear, any linear combination of such functions is still a solution. The linear combination that allows one to satisfy both the initial condition $P(x,N_\uin \vert x_\uin, N_\uin)=\delta(x-x_\uin)$ and the absorbing condition $P(0,N\vert x_\uin, N_\uin)=0$ is given by
\bea
P(x,N\vert x_\uin, N_\uin)=\left[ f_\mathrm{x_\uin,0}(x,N) - f_\mathrm{-x_\uin,0}(x,N) \right] \Theta(x)\, ,
\eea
where the Heaviside function restricts the solution to the inflating domain $x>0$. This solution is obtained by adding a (negative) image of the unconstrained solution mirrored through the absorbing boundary, which is often referred to as the ``method of images''. 

From the solution to the Fokker-Planck equation, one can compute the survival probability $S_{x_\uin}(N)$, \ie~the probability for the field to be still inflating at time $N$
\bea
\label{eq:survival:proba:flat:well}
S_{x_\uin}(N)=\int_0^\infty \dd x P(x,N\vert x_\uin, N_\uin) = \mathrm{erf}\left[\frac{x_\uin}{\sqrt{2 (N-N_\uin)}}\right] .
\eea
Alternatively, this can be written as the probability that, starting from $x_\uin$, the first-passage time through the end of inflation $x_\uend$ exceeds $N-N_\uin$, hence
\bea
\label{eq:survival:proba:flat:well:alternative}
S_{x_\uin}(N) = \int_{N-N_\uin}^\infty \Pfpt{x_\uin}(\mathcal{N})\dd\mathcal{N}\, .
\eea
By equating the two expressions above, and differentiating with respect to $N$, one obtains $\Pfpt{x_\uin}(\mathcal{N}) = - \left.\partial S_{x_\uin}/\partial N\right\vert_{N=N_\uin+\mathcal{N}}$, hence
\bea
\label{eq:Levy}
\Pfpt{x_\uin}(\mathcal{N})=\frac{x_\uin}{\sqrt{2\pi}\mathcal{N}^{3/2}}\exp\left(-\frac{x_\uin^2}{2\mathcal{N}}\right) .
\eea
This is called a L\'evy distribution. One can see that it features a very heavy tail, since $\Pfpt{x_\uin}(\mathcal{N})\propto \mathcal{N}^{-3/2}$ at large $\mathcal{N}$. This implies that the probability to realise large excursion times $\mathcal{N}$, hence large curvature perturbations according to \Sec{sec:SeparateUniverse}, is much larger than what would be obtained with Gaussian statistics. Although this conclusion is obtained with a toy, arguably unrealistic model, it is in fact entirely generic, as we shall now see.

\subsection{Adjoint Fokker-Planck equation}

In general, the survival probability starting from a given configuration $\vec{\Phi}_\uin$ can be defined in a way analogous to \Eq{eq:survival:proba:flat:well}, \ie
\bea
S_{\vec{\Phi}_\uin}(N) = \int_\Omega \dd\vec{\Phi} P\left(\vec{\Phi},N\vert\vec{\Phi}_\uin,N_\uin\right) .
\eea
Since $ P(\vec{\Phi},N\vert\vec{\Phi}_\uin,N_\uin)$ satisfies the adjoint Fokker-Planck equation~\eqref{eq:FP:adjoint}, so does the survival probability, $(\partial /\partial N_\uin )S_{\vec{\Phi}_\uin}(N)  = \mathcal{L}^\dagger(\vec{\Phi}_\uin)\cdot S_{\vec{\Phi}_\uin}(N) $. Since $\vec{F}(\vec{\Phi})$ and $\vec{G}(\vec{\Phi})$ do not depend on time explicitly, the survival probability only depends on the time difference $N-N_\uin$, hence $(\partial /\partial N_\uin )S_{\vec{\Phi}_\uin}(N)  = - (\partial /\partial N )S_{\vec{\Phi}_\uin}(N) $. Moreover, similarly to \Eq{eq:survival:proba:flat:well:alternative} one can write $S_{\vec{\Phi}_\uin}(N) = \int_{N-N_\uin}^\infty \Pfpt{\vec{\Phi}_\uin}(\mathcal{N})\dd\mathcal{N}$. Combining these results, one obtains
\bea
\label{eq:FP:adj:Pfpt}
\frac{\partial}{\partial \mathcal{N}} \Pfpt{\vec{\Phi}_\uin}(\mathcal{N}) = \mathcal{L}^\dagger(\vec{\Phi}_\uin)\cdot\Pfpt{\vec{\Phi}_\uin}(\mathcal{N}) \, ,
\eea
\ie~the first-passage time distribution obeys the adjoint Fokker-Planck equation. The only difference with the Kolmogorov backward equation~\eqref{eq:FP:adjoint} is the boundary conditions: here, \Eq{eq:FP:adj:Pfpt} needs to be solved with $ \Pfpt{\vec{\Phi}_\uin}(\mathcal{N})=\delta(\mathcal{N})$ when $\vec{\Phi}_\uin\in\mathcal{C}_\uend$.
In the free-diffusion problem, one can check that, indeed, the L\'evy distribution \eqref{eq:Levy} satisfies the adjoint Fokker-Planck equation, which in that case coincides with the Fokker-Planck equation~\eqref{eq:free:diff:FP} since $F=0$ and $G$ is uniform.

\section{Heavy tails and non-perturbative non-Gaussianities}\label{sec:NonPertNG}

The first-passage-time distribution depends on the details of the inflationary model under consideration, and a variety of profiles can be obtained, see for instance Refs.~\cite{Pattison:2017mbe, Panagopoulos:2019ail, Figueroa:2020jkf, Pattison:2021oen, Ezquiaga:2019ftu, Vennin:2020kng, Achucarro:2021pdh, Kitajima:2021fpq, Cai:2022erk, Animali:2022otk, Jackson:2022unc, Ezquiaga:2022qpw, Rigopoulos:2022gso, Briaud:2023eae, Hooshangi:2023kss, Kawaguchi:2023mgk}. In all cases however, the tail of the distribution, which is relevant for the formation of PBHs, is heavier than the Gaussian behaviour predicted by linear perturbation theory. In this section, we briefly review why.

\subsection{Characteristic function}

The adjoint Fokker-Planck equation~\eqref{eq:FP:adj:Pfpt} is a linear partial differential equation, hence it can be turned into a set of ordinary differential equations by introducing the characteristic function
\bea\label{eq:def:chi:Fourier}
\chi\left(t,\vec{\phi}\right)=\left\langle e^{it\mathcal{N}} \right\rangle = \int_{-\infty}^{\infty} \dd \mathcal{N} e^{i t \mathcal{N}} \Pfpt{\vec{\Phi}}(\mathcal{N})\, .
\eea
This is nothing but the Fourier transform of the first-passage-time distribution, which is a function of the dummy parameter $t$. Using \Eq{eq:FP:adj:Pfpt}, it satisfies the differential equation 
\bea
\label{eq:eom:chi}
\mathcal{L}^\dagger\left(\vec{\Phi}\right)\cdot \chi(t,\vec{\Phi}) = -i t \chi\left(t,\vec{\Phi}\right) ,
\eea
together with the boundary condition $\chi(t,\vec{\Phi})=1$ when $\vec{\Phi}\in\mathcal{C}_\uend$. 

\subsection{Pole expansion}

In general, the presence of non-trivial boundary conditions results in the existence of poles, and the characteristic function can be expanded as
\bea
\label{eq:pole:expansion}
 \chi\left(t,\vec{\Phi}\right) =  f\left(t,\vec{\Phi}\right) + \sum_n \frac{a_n(\vec{\Phi})}{\Lambda_n-i t}
\eea
where $f$ is a regular function, and we have assumed that the poles form a discrete set. The form~\eqref{eq:pole:expansion} follows from the structure of the Fokker-Planck operator and can be justified formally, see Ref.~\cite{Ezquiaga:2019ftu}, where the parameters $\Lambda_n$ are shown to correspond to the eigenvalues of the adjoint Fokker-Planck operator, hence they do not depend on $\vec{\Phi}$. 

Let us illustrate this decomposition with the example of free diffusion studied in \Sec{eq:free:diffusion}, where an additional reflective boundary is added at $\phi_{\mathrm{r}}$. Such a model is referred to as a ``flat well''. This boundary can be removed by evaluating the formulas below in the limit $\phi_{\mathrm{r}}\to \infty$ but at this stage $\phi_{\mathrm{r}}$ needs to be kept finite, since only when adding the boundary condition $(\partial/\partial \phi) \chi(t,\phi)\vert_{\phi=\phi_{\mathrm{r}}}=0$ is the solution to \Eq{eq:eom:chi} fully specified, and reads
\bea
\label{eq:chi:flat:well}
\chi(t,\phi) = \frac{\cos\left[\sqrt{it}\mu\left(\frac{\phi}{\phi_{\mathrm{r}}}-1\right)\right]}{\cos\left(\sqrt{it}\mu\right)}
\quad\quad \text{where} \quad\quad \mu^2=\frac{24\pi^2\Mp^2\phi^2_{\mathrm{r}}}{V}\, 
\eea
and where we have set $\phi_\uend=0$ without loss of generality.
The partial fraction decomposition of $\chi(t,\phi)$ is of the form~\eqref{eq:pole:expansion}, with $\Lambda_n=(\pi/\mu)^2(n+1/2)^2$ and $a_n(\phi)=(-1)^n \pi (2n+1)\cos[\frac{\pi}{2}(2n+1)(\frac{\phi}{\phi_{\mathrm{r}}}-1)]/ \mu^2$. 

The inverse Fourier transform that is required to derive the first-passage-time distribution can be performed by means of the residue theorem, and one obtains
\bea
\label{eq:Pfpt:exp:expansion}
\Pfpt{\vec{\Phi}}\left(\mathcal{N}\right) = \sum_n a_n(\vec{\Phi}) e^{-\Lambda_n \mathcal{N}}\, .
\eea
On the tail, \ie~at large values of $\mathcal{N}$, the lowest eigenvalues $\Lambda_n$ dominate, hence if the poles are ordered such that $\Lambda_0<\Lambda_1<\cdots$, the pole expansion is essentially a tail expansion. 

\subsection{Tail behaviour}
\label{sec:tail}

 \begin{figure}
\centering
\includegraphics[scale=0.6]{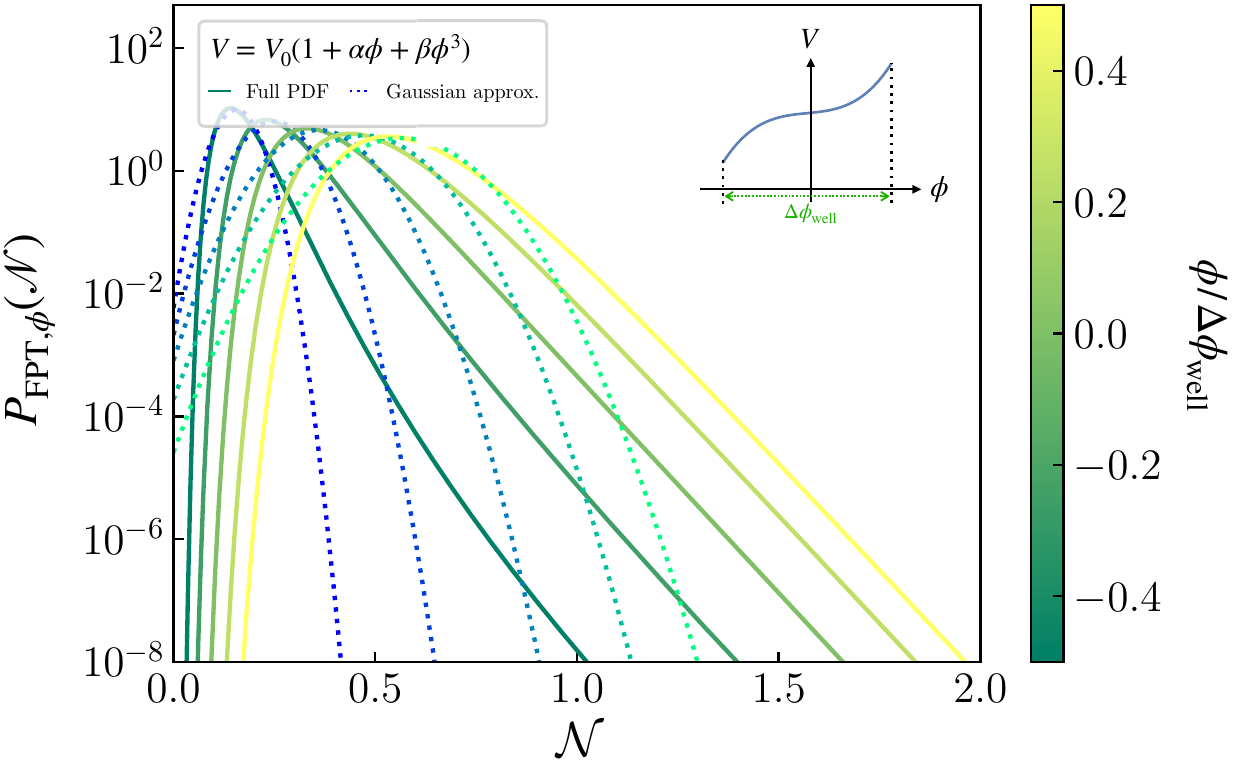}
\caption{Probability density functions of the number of \efolds~generated by an inflection-point inflationary potential (see inset, with $V_0/(24\pi^2\Mp^4)=10^{-3}$, $\alpha=0.24/\Mp$, $\beta=9\Mp^3$ and $\Delta\phi_{\mathrm{well}}=5.83\Mp$ -- the value for $V_0$ is unrealistically large but it is used for convenience of the illustration). Solid lines correspond to the full distribution functions computed by mean of the stochastic-$\delta N$ formalism, where different colours correspond to different initial conditions. The dotted lines correspond to the Gaussian approximation provided by linear perturbation theory, which provides a good fit for the maximum of the distribution but however fails to describe the heavy tails, where PBHs are nonetheless produced. 
Adapted from Ref.~\cite{Ezquiaga:2019ftu}.} \label{fig:Pfpt}
\end{figure}

In Fig.~\ref{fig:Pfpt}, we display the first-passage-time distribution obtained in a tilted-inflection point model, for a few initial conditions. The solid lines correspond to the numerical solutions of \Eq{eq:FP:adj:Pfpt} (in practice, we solved \Eq{eq:eom:chi} on a grid of $t$ values and inverse-Fourier transformed the result), while the dotted lines correspond to the prediction of linear cosmological perturbation theory. In perturbation theory, one obtains Gaussian distributions, which provide good fits to the full distributions close to the maximum. This is because the statistics of the most likely fluctuations are almost Gaussian. However, in the tail of distribution, the Gaussian approximation breaks down. Indeed, when $\mathcal{N}$ becomes large, \Eq{eq:Pfpt:exp:expansion} reduces to $\Pfpt{\vec{\Phi}}\left(\mathcal{N}\right) \simeq a_0(\phi) e^{-\Lambda_0 \mathcal{N}}$, \ie~the tail has an exponential profile. Compared to Gaussian suppression, $\Pfpt{\vec{\Phi}}\left(\mathcal{N}\right) \propto e^{- \propto \mathcal{N}^2}$, the tail is much heavier, and since large values of $\mathcal{N}$ correspond to large values of $\zeta$, this can have a strong impact on the abundance of PBHs~\cite{Gow:2022jfb}. 

If $\zeta$ has Gaussian statistics, the probability to realise $\zeta>\zeta_\uc\sim 1$ is of order $p_\uc=\mathrm{erfc}[\zeta_\uc/(\sqrt{2 \mathcal{P}_\zeta})]/2 \propto e^{-\zeta_\uc^2/(2 \mathcal{P}_\zeta)}$, where $\mathrm{erfc}$ is the complementary error function, which we have expanded in the regime $\zeta_\uc\gg\sqrt{\mathcal{P}_\zeta}$. PBH dark-matter typically requires the PBH mass fraction to be of order $p_\uc\sim 10^{-20}$ at formation, which thus corresponds to $\mathcal{P}_\zeta\simeq 10^{-2}$. For this reason it is generally considered that, for dark matter to be comprised of PBHs, such large values of the power spectrum are required, which can be challenging to achieve from a model-building perspective. However, in the presence of exponential tails, the situation is quite different. For instance, consider the case where the PDF of $\zeta$ is Gaussian across $n_{\mathrm{SD}}$ standard deviations around its maximum, and then exponential. In that case $p_\uc \simeq e^{-n_{\mathrm{SD}}^2/2-\Lambda_0}/(\Lambda_0 \sqrt{2\pi}\mathcal{P}_\zeta)$, which clearly signals that cosmologically-relevant values of $p_\uc$ do not require large values of the power-spectrum. This is why it is crucial to take quantum diffusion into account when it comes to computing the abundance and statistics of PBHs.

Finally, let us note that a compact inflating domain is required for the eigenvalues $\Lambda_n$ to form a discrete set. Indeed, separated poles in the characteristic function follow from imposing boundary conditions at finite field distances, in the same way that standing waves can assume only certain frequencies in a finite-sized cavity. If the cavity becomes infinite, any frequency is allowed, and likewise, if the inflating domain is not bounded, $n$ becomes a continuous rather than discrete index in \Eq{eq:Pfpt:exp:expansion}. An explicit example of this phenomenon is given by the pure-diffusion example: in that case, $\Lambda_{n+1}-\Lambda_n = 2(n+1)\pi^2/\mu^2$, which goes to zero, since $\mu^2\to\infty$, when $\phi_{r}\to\infty$, hence the set of poles becomes continuous. When this occurs the tail of $\Pfpt{\vec{\Phi}}(\mathcal{N})$ can no longer be approximated by $a_0(\vec{\Phi})e^{-\Lambda_0\mathcal{N}}$, hence it becomes even heavier than exponential. In the free-diffusion example, in the limit $\phi_{\mathrm{r}}\to \infty$ we  found a L\'evy distribution, see \Eq{eq:Levy}, for which the tail is indeed monomial rather than exponential. This makes the conclusions derived above even more relevant in that case.

\subsection{Non-perturbative non-Gaussianities}
\label{sec:Non:Pert:NG}

 \begin{figure}
\centering
\includegraphics[scale=0.6]{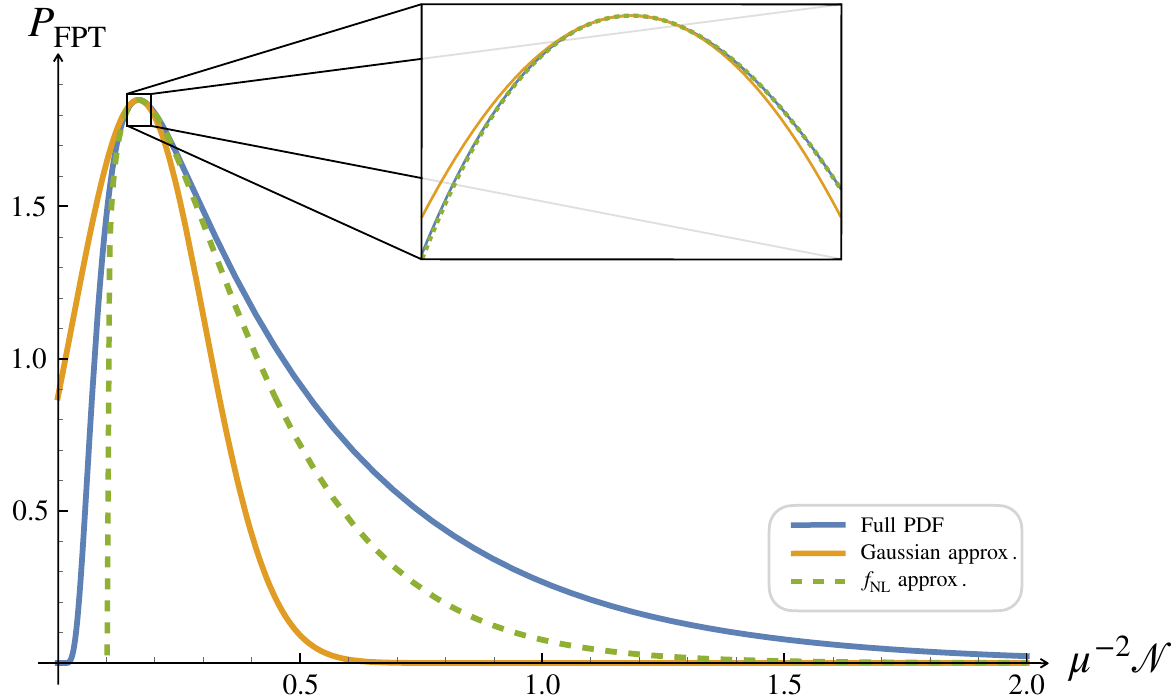}
\caption{Probability density function of the number of \efolds~generated by a flat well starting from $\phi=\phi_{\mathrm{r}}$, see \Eqs{eq:chi:flat:well}-\eqref{eq:Pfpt:exp:expansion}. The orange line corresponds to the Gaussian approximation, while the dashed green line stand for the local $\fNL$ expansion, see \Eq{eq:local:fNL}. The inset zooms in on the maximum of the distribution, where the Gaussian and quasi-Gaussian approximations provide good fits. However, they both fail to properly describe the heavy tails.} \label{fig:free:diffusion}
\end{figure}

The importance of non-Gaussianities in shaping the detailed PBH statistics is well-known, but so far it has been mostly investigated using perturbative techniques, such as $\fNL$-type parametrisations. These expansions are perturbative in nature, \ie~they expand around Gaussian statistics, introducing corrections arising from the bispectrum (the three-point correlation function), the trispectrum (the four-point correlation function), etc. For instance, if $\mathcal{N}_\mathrm{G}$ is a Gaussian random variable, the local-$\fNL$ parametrisation corresponds to the approximation
\bea
\label{eq:local:fNL}
\mathcal{N} \simeq \mathcal{N}_{\mathrm{G}} + \fNL \left( \mathcal{N}_{\mathrm{G}}^2 - \left\langle  \mathcal{N}_{\mathrm{G}} ^2 \right\rangle \right) .
\eea
The parameter $\fNL$ can be extracted from the third moment of $\mathcal{N}$, which consists of  Taylor expanding the PDF of $\mathcal{N}$ around its maximum. This approximation is compared to the full distribution in Fig.~\ref{fig:free:diffusion} in the free-diffusion toy model. This illustrates that, although the $\fNL$ approximation improves the quality of the fit close to the maximum, it fails to capture the tail behaviour, which is crucial for the PBH abundance.

The reason for this failure is that the non-Gaussianities produced by quantum diffusion are non-perturbative in nature. Thus, they require non-perturbative parametrisations. A straightforward non-perturbative generalisation of \Eq{eq:local:fNL} consists of treating the local-$\fNL$ parametrisation as a Taylor expansion of an underlying local ansatz $\mathcal{N}(\vec{x})=f[ \mathcal{N}_{\mathrm{G}}(\vec{x}),\vec{x} ]$, where $f$ is any (in general non-linear) function. This approach was recently followed in Refs.~\cite{Kitajima:2021fpq, Gow:2022jfb}. It has the advantage of allowing one to derive the statistics of the compaction function, hence the mass fraction of PBHs, in a consistent non-perturbative way, which is otherwise difficult (see \Sec{sec:from:zeta:to:C} below). However, it is not yet clear whether or not non-Gaussianities inherited from quantum diffusion are well-approximated by the above local ansatz, and further investigations are required.

\section{Prospects}\label{sec:Prospects}

Let us end this chapter by reviewing a few of the open challenges that remain to be addressed to better understand the role of quantum diffusion in shaping the statistics of PBHs. Our goal is not to list all of them, but simply to highlight a few conceptual and technical issues that we think deserve further attention in the years to come.

\subsection{The stochastic-$\delta N$ formalism beyond slow roll}

Most applications of the stochastic-$\delta N$ formalism to-date have been restricted to slow-roll inflation in which the coarse-grained fields follow an attractor trajectory where the fields' momenta are uniquely determined by the field values, reducing the phase-space to 1D for a single scalar field, for example. However models of inflation where there is a significant enhancement of primordial density perturbations on small scales necessarily break the single-field and/or slow-roll assumptions and include transient, non-attractor behaviour, such as ultra-slow roll. 

Extending the phase-space beyond the slow-roll trajectory has several important consequences.
Firstly, and most obviously, it raises the dimensionality of the probability distribution function $P(\vec{\Phi},N \vert \vec{\Phi}_\uin,N_\uin)$ and the Fokker-Planck equation~(\ref{eq:Fokker:Planck}), and hence makes more challenging the first-passage time problem to find $\Pfpt{\vec{\Phi}_\uin}(\mathcal{N})$. 
In the case of a single field, for example, extending the phase space beyond slow roll~\cite{Grain:2017dqa} leads to the the characteristic function (\ref{eq:def:chi:Fourier}) obeying a partial differential equation~\cite{Pattison:2021oen} rather than an ordinary differential equation as in the case of slow roll for a single field~\cite{Pattison:2017mbe,Ezquiaga:2019ftu}. 
The first-passage-time problem in 2D has been studied in the ultra-slow roll limit, assuming the usual Bunch-Davies vacuum state for a massless field in de Sitter~\cite{Firouzjahi:2018vet,Pattison:2021oen}. In the small-velocity (diffusion-dominated) expansion one recovers at leading order the previous slow-roll results~\cite{Pattison:2017mbe}, but at next-to-leading order we find an additional series of poles in the characteristic function. This does not affect the lowest pole, and hence the leading exponential term in the tail of the probability distribution for the first passage time, but it can suppress the amplitude of the tail.

There are several other challenges. 
As we move away from the massless limit the power spectrum for the stochastic noise due to quantum fluctuations crossing outside the coarse-graining scale deviates from the simple result for a massless field in de Sitter, $\langle\xi_\varphi^2\rangle=(H/2\pi)^2$, and one must also include the noise affecting the field momentum, $\xi_\pi$~\cite{Grain:2017dqa,Figueroa:2020jkf,Cruces:2021iwq,Mishra:2023lhe,Iacconi:2023slv}. In slow roll, the fluctuations are overdamped on super-horizon scales leading to a squeezed state that can be treated as classical noise, but in the absence of an overdamped attractor then the question arises as to whether a classical treatment remains sufficient.
Moreover, away from slow roll the covariance of the noise does not only depend on the instantaneous value of $H$, but it is also affected by the details of the prior background evolution. This implies that the field mode equation should be solved along each realisation of the Langevin equation, giving rise to non-Markovian processes that are challenging to investigate~\cite{Figueroa:2020jkf,Figueroa:2021zah}.

Finally, we note that single-field models leading to a rapid rise in the primordial power spectrum on small scales require a sudden, non-adiabatic transition to force the background field out of slow-roll into an ultra-slow-roll phase. The sudden transition leads to deviations from the Bunch-Davies vacuum state (\ie~particle production) on sub-Hubble scales, and non-adiabatic perturbations, sourced by residual spatial gradients, on a finite range of super-Hubble scales at the transition~\cite{Leach:2001zf}. These gradient terms are not included in the usual separate-universe approximation and require a re-evaluation of the classical and stochastic $\delta N$ formalisms at a sudden transition~\cite{Jackson:2023obv}.

\subsection{From the curvature perturbation to the compaction function}
\label{sec:from:zeta:to:C}

So far in this chapter, the production of PBHs has been attributed to large local values of $\zeta_R$, the curvature perturbation coarse-grained at the scale $R$, where the mass of the resulting black hole scales with the Hubble mass at the time the scale $R$ re-enters the Hubble radius. This is however only an approximation, and modern PBH formation criteria rather depend on the smoothed density contrast or the compaction function~\cite{Shibata:1999zs, Harada:2015yda, Musco:2018rwt} to determine whether a PBH forms, and use critical scaling to infer its mass~\cite{Choptuik:1992jv, Evans:1994pj, Koike:1995jm}.

The curvature perturbation is a direct outcome of the stochastic $\delta N$ formalism but the statistics of other relevant cosmological fields, at the non-perturbative level, remain unexplored. One reason why the curvature perturbation is not ideal to predict the formation of PBHs at the scale $R$ is that it is sensitive to power at larger scales. Indeed, if $\zeta_R$ denotes the curvature perturbation coarse-grained at the scale $R$, from the coincident two-point function, $\langle \zeta_R^2(\vec{x})\rangle = \int \dd\ln k \mathcal{P}_\zeta(k) \mathrm{sinc}(kR/a)$, one can see that power at scales $k\ll a/R$ may contribute significantly to $\zeta_R$, depending on the power spectrum. However, at large scales, $\zeta$ is conserved and may be interpreted as a renormalisation of the scale factor seen by a local observer, hence it should not affect the collapse dynamics of a local over-density. 
In contrast, at the linear level the comoving density contrast is related to the curvature perturbation via 
\bea
\delta_{\mathrm{lin}} = \frac{2(1+w)}{5+3w} \frac{k^2}{a^2 H^2} \zeta\, ,
\eea
where $w$ is the equation-of-state parameter. The density contrast thus features an additional $k^2$ suppression at large scales, which ensures that only the scales $k \sim a/R$ contribute to the formation of black holes with size $R$. If one accounts for the non-linear relationship between the curvature perturbation and the density contrast, one is specifically led to consider the compaction function, but the argument remains the same. 

One possible approach to compute the statistics of the density contrast or the compaction function is to consider the coarse-shelled curvature perturbation, where instead of averaging $\zeta$ over a sphere of radius $R$, one considers its average between two spheres of radii $R_1$ and $R_2$, $\Delta\zeta_{R_1,R_2}(\vec{x}) = \zeta_{R_2}(\vec{x}) - \zeta_{R_1}(\vec{x})$. This selects scales in the interval $a/R_2<k<a/R_1$, so by adjusting the parameters $R_1$ and $R_2$ around $R$, one can try to mimic $\delta_R(\vec{x})$. Since the coarse-shelled curvature perturbation can be written as the difference between two coarse-grained curvature perturbations, its statistics can be computed in the stochastic-$\delta N$ formalism, and in this way one may approximate the statistics of the comoving density contrast. This approach is developed in Ref.~\cite{Tada:2021zzj}, where it is also applied to the compaction function. This requires us to compute the two-point distribution of the curvature perturbation, $P[\zeta_{R_1}(\vec{x}),\zeta_{R_2}(\vec{x})]$, but no higher statistics.

In those rare cases where all $n$-point functions are known, exact calculations can also be performed. For instance, under the generalised local ansatz mentioned in \Sec{sec:Non:Pert:NG}, $\zeta(\vec{x}) = f[\zeta_{\mathrm{G}}(\vec{x}),\vec{x}]$ where $\zeta_{\mathrm{G}}$ is a Gaussian field, the statistics of the compaction function $\mathcal{C}_R$ can be derived using 
\bea
\mathcal{C}_R=\frac{2}{3}\left[1-\left(1+R\zeta_R'\right)^2\right]\, ,
\eea
where a prime denotes differentiation with respect to $R$. This approach was followed in Refs.~\cite{Kitajima:2021fpq, Gow:2022jfb}, where the importance of heavy tails was confirmed. Similarly, in the setup recently considered in Ref.~\cite{Raatikainen:2023bzk}, an explicit local expression is given for $\zeta_R(\vec{x})$ in terms of a set of Gaussian fields, which makes the computation of the compaction function tractable. 

Beyond those simple cases however, the derivation of the compaction function in the stochastic-$\delta N$ formalism remains an open issue. The use of more advanced structure-formation tools, such as the excursion set~\cite{Peacock:1990zz, Bower:1991kf, Bond:1990iw} or peak theory~\cite{Bardeen:1985tr}, together with critical scaling, also remains to be explored.

\subsection{Volume weighting and backward observables}

In the standard picture of cosmological perturbation theory, along a given field-phase space classical trajectory, there is a one-to-one relationship between comoving scales $k$ and the system's configuration $\vec{\Phi}_*(k)$ when that scale crosses out the Hubble radius during inflation. Therefore, by performing measurements of the statistics at the scale ${k}$, one is able to probe the inflationary setup (the potential function, the field geometry, etc) only around that location $\vec{\Phi}_*$. In the stochastic picture however, such a one-to-one relationship does not exist. Assuming that $H$ is quasi-constant during inflation, a given scale $k$ exits the Hubble radius at the time $N_*(k)=N_\uend-\ln(k_\uend/k)$, where $k_\uend$ denotes the comoving Hubble scale at the end of inflation. Therefore, if $\vec{\Phi}_*$ denotes the system's configuration at the time $N_*$, then it is a random quantity, and its distribution function is given by~\cite{Ando:2020fjm}
\bea
P_{\mathrm{bw}}\left(\vec{\Phi}_*\right)=\Pfpt{\vec{\Phi}_*}\left(N_{\mathrm{bw}}\right) \frac{\int_0^\infty P\left(\vec{\Phi}_*,N \vert \vec{\Phi}_\uin, N_\uin\right) \dd N }{ \int_{N_{\mathrm{bw}}}^\infty  \Pfpt{\vec{\Phi}_\uin}\left(N_{\mathrm{tot}}\right) \dd N_{\mathrm{tot}}}\, .
\eea
Here, $N_{\mathrm{bw}} = \ln(k_\uend/k)$ is the number of $e$-folds corresponding to the Hubble exit of $k$, counted backwards from the end of inflation. When the model under consideration features regions with large quantum diffusion, this distribution function may have a wide support, which implies that a broad region of field-phase space can be probed even if a limited range of scales is observationally accessible. Note that this is the case even if the scales being measured emerge prior to the diffusion-dominated region, which allows one to place constraints on PBH-compliant models from CMB measurements only~\cite{Ando:2020fjm}.

 \begin{figure}
\centering
\includegraphics[scale=0.4]{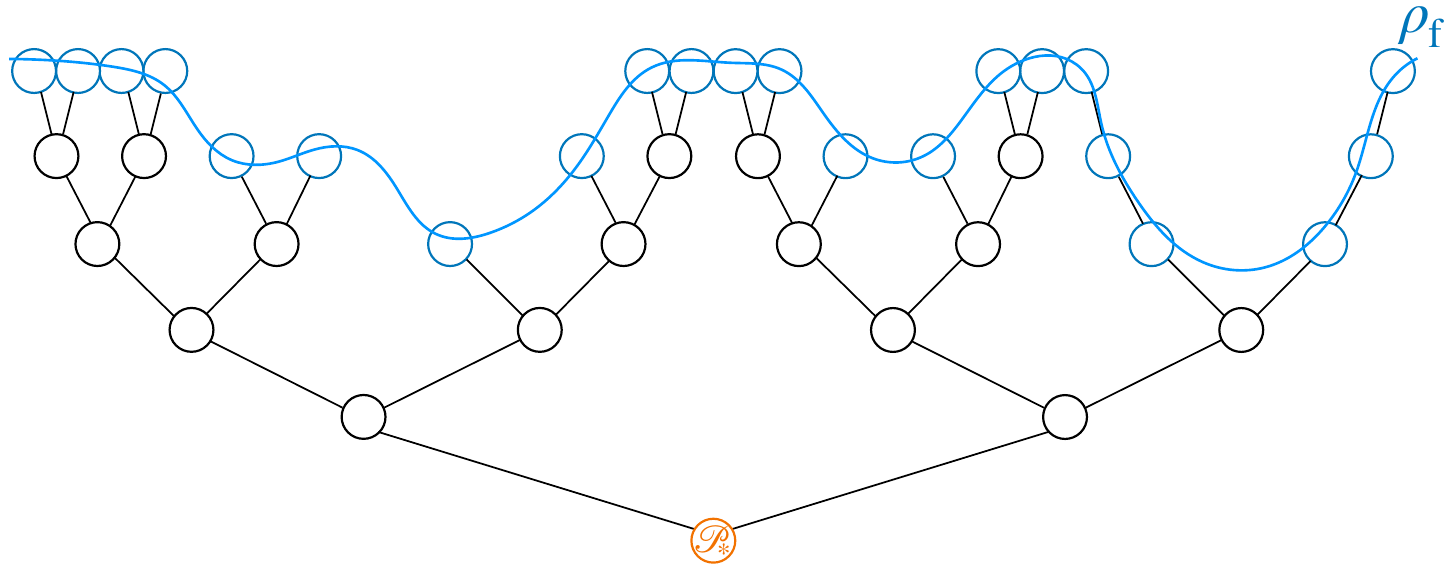}
\caption{Space-time structure of a stochastically-inflating universe (adapted from Ref.~\cite{Animali:2024jiz}).} \label{fig:BinaryTree}
\end{figure}

In practice, the relationship between field values and physical distances is even more involved, and is encoded in the recursive geometry of a stochastically inflating universe. Consider indeed the situation sketched in Fig.~\ref{fig:BinaryTree}. Here, each disk stands for a Hubble patch, and as inflation proceeds each patch gives rise to new Hubble patches, until the end-of-inflation hypersurface $\mathcal{C}_\uend$ is being crossed. In this picture, the inflating space-time has the structure of a tree in the language of graph theory. From a given initial patch, denoted $\mathcal{P}_*$ in the figure and with field value $\vec{\Phi}_*$, a certain volume $V$ is produced on the final hypersurface, which corresponds to the number of leaves originating from the root $\mathcal{P}_*$. The distribution function $P(V\vert\vec{\Phi}_*)$ is not straightforward to compute, since all branches originating from the root $\mathcal{P}_*$ coincide until they split, hence the leaves cannot be seen as the end-point of independent stochastic realisations of the Langevin equation. Methods can nonetheless be developed to reconstruct the distribution of the final volume~\cite{Animali:2024jiz}, which contains the relevant information to match distances as measured by local observers on the final hypersurface to field configurations when the corresponding spatial region emerges during inflation. 

Subtleties may however arise when considering volume weighting. Consider indeed a local observer on the final hypersurface of the tree in Fig.~\ref{fig:BinaryTree}, and task them with measuring the mean density on that hypersurface. What they would do is to measure the density in each leaf, and compute the ensemble average. As mentioned above, ensemble averages over subsets of leaves correspond to volume averages from the viewpoint of the root patch $\mathcal{P}_*$. The reason is that regions of the universe that inflate more give rise to a larger number of leaves on the final hypersurface, hence they contribute more to ensemble averages computed by local observers. As a consequence, when it comes to observable quantities, distributions should be volume-weighted by $e^{3\mathcal{N}}$. For instance, for the curvature perturbation coarse-grained at $R_\uend$, the Hubble radius at the end of inflation, one has
\bea
P\left(\zeta_{R_\uend}\right) \propto \Pfpt{\vec{\Phi}_\uin}\left(\left\langle \mathcal{N}_{\vec{\Phi}_\uin}\right\rangle_V+\zeta_{R_\uend}\right) e^{3 \zeta_{R_\uend}}\, ,
\eea
where the overall factor is set such that the distribution is normalised, and $\langle \cdot\rangle_V$ is the volume-weighted stochastic average, i.e.
\bea
\left\langle \mathcal{N}_{\vec{\Phi}_\uin}\right\rangle_V = \frac{\int \Pfpt{\vec{\Phi}_\uin}(\mathcal{N}) e^{3\mathcal{N}} \mathcal{N} \dd \mathcal{N}}{\int \Pfpt{\vec{\Phi}_\uin}(\mathcal{N}) e^{3\mathcal{N}}  \dd \mathcal{N}}\, .
\eea
This is what a local observer would measure by recording $\mathcal{N}_{\vec{\Phi}_\uin}$ in each leaf and computing the ensemble average. Recalling that the first-passage-time distribution has an exponential tail, $\Pfpt{\vec{\Phi}_\uin}(\mathcal{N})\propto e^{-\Lambda_0 \mathcal{N}}$, the above formula fails to converge when $\Lambda_0\leq 3$. This is the case where ``eternal inflation'' takes place, and the mean volume originating from a given patch $\mathcal{P}_*$ diverges. Note that this issue cannot be avoided by starting inflation at sufficiently low energy: as mentioned in \Sec{sec:tail}, $\Lambda_0$ is independent of $\vec{\Phi}$, hence if volume divergences arise somewhere then they arise everywhere. Note also that, as mentioned in \Sec{sec:tail}, when the field-phase space is not compact, the tail of the first-passage time distribution is even heavier than exponential, hence volume divergences necessarily occur. This is the case for most inflationary models considered in the literature, hence deriving properly volume-weighted observables in this context remains an important challenge to be addressed.

\section*{Acknowledgement}
We would like to thank all the colleagues with whom we have investigated quantum diffusion and primordial black holes over the past few years, in particular Kenta Ando, Chiara Animali, Hooshyar Assadullahi, Vadim Briaud, Jose-Mar\'ia Ezquiaga, Juan Garc\'ia-Bellido, Andrew Gow, Joseph Jackson, Kazuya Koyama, Chris Pattison, Yuichiro Tada, Koki Tokeshi and Eemeli Tomberg.

\bibliographystyle{JHEP}
\bibliography{biblio}

\end{document}